\documentclass[12pt]{article}

\usepackage{moreverb}

\usepackage[colorlinks,bookmarksopen,bookmarksnumbered,citecolor=red,urlcolor=red]{hyperref}

\newcommand\BibTeX{{\rmfamily B\kern-.05em \textsc{i\kern-.025em b}\kern-.08em
T\kern-.1667em\lower.7ex\hbox{E}\kern-.125emX}}

\usepackage{graphicx}
\usepackage{natbib}
\usepackage{lscape,rotating}
\usepackage{float,epsfig}
\usepackage{subfigure}
\usepackage{setspace}
 \usepackage{amsmath,bm} 
\usepackage{amsfonts}
\usepackage{amssymb}

\newcommand{\xv}{\bm{x}}

\newcommand{\zv}{\bm{z}}

\newcommand{\betav}{\bm{\beta}}

\begin{document}

\def\spacingset#1{\renewcommand{\baselinestretch}%
{#1}\small\normalsize} \spacingset{1}
\setlength{\parindent}{3ex}
\spacingset{1.25} 

\title {\bf Notes on Exact Power Calculations for t Tests and Analysis of Covariance}
\author{Yongqiang Tang \footnote{email: yongqiang\_tang@yahoo.com} \\
    Tesaro, 1000 Winter St, Waltham, MA 02451 }
\maketitle
\abstract{ \cite{tang:2018a, tang:2018e}
 derived the exact power formulae for t tests and  analysis of covariance (ANCOVA) in superiority, noninferiority and equivalence  trials. The power calculation in equivalence trials can be simplified by using Owen's Q function, which is available in standard statistical software. We
 extend the exact power determination method for ANCOVA to unstratified and stratified multi-arm randomized trials. The method is applied to the design of multi-arm trials and  gold standard noninferiority  trials }

\noindent%
\bigskip

\section{Introduction}

 \cite{tang:2018e, tang:2018a} obtained the exact power formulae for some commonly used t tests  in superiority, noninferiority (NI) and equivalence  trials. 
The  power determination for the analysis of covariance (ANCOVA) and t-test with unequal variances in equivalence trials  involves two-dimensional numerical integration.
We show that  the calculation can be simplified by using Owen's Q function, which is available in  standard statistical software packages (e.g. SAS and R {\it PowerTOST }). 
 We   extend the  method for ANCOVA to  unstratified and stratified  multi-arm randomized trials, and apply it to 
the power determination for  multi-arm trials and gold standard NI trials \citep{pigeot:2003}.

We use the same notations as  \cite{tang:2018e, tang:2018a}. Let $t(f,\lambda)$ denote the t distribution with $f$ degrees of freedom and  noncentrality parameter $\lambda$, 
 $t_{f,p}$ the $p$th percentile of the  central  t distribution, $\Phi(\cdot)$  the cumulative distribution function (CDF)  of $N(0,1)$, $F_{f_1,f_2}(\cdot)$ the CDF of a central $F(f_1,f_2)$ distribution, and
$Q_f(t,\delta;a,b)=\frac{1}{\Gamma(f/2)2^{f/2-1}}\int_a^b \Phi(\frac{tx}{\sqrt{f}}-\delta)x^{f-1}\exp(-\frac{x^2}{2})dx$  Owen's Q function.
Let $n_g$ be the number of subjects in group $g$, $n$ the total size, $M_0$ the superiority ($M_0=0$) or NI margin, and $(M_l,M_u)$ the lower and upper equivalence margins.
Without loss of generality, we assume high scores indicate better health.

\section{Two sample t tests}
Let  $(\hat\tau, n^{-1}\hat{V})$ be the estimated effect and variance with true values  $(\tau_1, n^{-1}V)$ in a test based on the t distribution.
Suppose $\frac{\hat\tau-\tau_1}{\sqrt{n^{-1}V}}\sim N(0,1)$ is independent of $\xi= \frac{\hat{V}}{{V}} \sim \frac{\chi_f^2}{f}$. 
In superiority and NI trials, we reject the null hypothesis when $t=\frac{\hat\tau -M_0}{\sqrt{n^{-1}\hat{V}}} >C= t_{f,1-\alpha/2}$. 
If  $f$ and $V$ are known,
the exact power   is $\Pr\left[t(f, \frac{|\tau_1-M_0|}{\sqrt{n^{-1}V}})> C\right]$, or  $1$ minus the CDF of  $t \sim t(f, \frac{|\tau_1-M_0|}{\sqrt{n^{-1}V}})$ evaluated at $C$.

An equivalence test is  significant if both $t_l=\frac{\hat\tau -M_l}{\sqrt{n^{-1}\hat{V}}} > C$ and $t_u=\frac{\hat\tau -M_u}{\sqrt{n^{-1}\hat{V}}} <-C$.  By the change of variable
$x=\sqrt{f\xi}$,  the  exact power equation (26) of \cite{tang:2018a} can be rearranged   in terms of Owen's Q function as
\begin{equation}\label{power00equi0}
 P_{\text{equi}} =\int_0^{ \frac{(M_u-M_l)^2}{4 n^{-1}V C^2}} \left[\Phi(\delta_1-  C \sqrt{\xi}) -    \Phi(\delta_2+  C\sqrt{\xi}) \right]  dG (\xi)
={\it Q_f(-C,\delta_2; 0,R)- Q_f(C, \delta_1;0,R)}\end{equation}
where    $G(\xi)$ is the CDF of $\xi\sim \frac{\chi_f^2}{f}$, $\delta_2=\frac{M_l-\tau_1}{\sqrt{ n^{-1}V}}<0$, $\delta_1=\frac{M_u-\tau_1}{ \sqrt{n^{-1}V}}>0$ and  $R=\frac{\sqrt{f}(\delta_1-\delta_2)}{2 C}$.

In the t test with unequal variances [i.e. $y_{0i}\stackrel{iid}{\sim} N(\mu_0,\sigma_0^2)$, $y_{1i}\stackrel{iid}{\sim} N(\mu_1,\sigma_1^2)$], the power of the superiority and NI trial is obtained from the fact \citep{moser:1989,tang:2018a} 
 that $ \frac{\hat\tau -M_0}{\sqrt{n^{-1}\hat{V}}} h^*(u) = \frac{\hat\tau-M_0}{\sqrt{n^{-1}V}}\sqrt{\frac{n-2}{(n_1-1)s_1^2/\sigma_1^2 +(n_0-1)s_0^2/\sigma_0^2}}$ follows a noncentral $ t(n-2,\frac{|\tau_1-M_0|}{\sqrt{ n^{-1} V}})$ distribution  given $u$ 
\begin{eqnarray}\label{powertsw_equi}
\begin{aligned}
P_{\text{sup/ni}} =\int_0^\infty\, \text{Pr}\left[ t\left(n-2,\frac{|\tau_1-M_0|}{\sqrt{ n^{-1}V}}\,\right) >h(u) \right]\,dF_{n_1-1,n_0-1}(u)
\end{aligned}
\end{eqnarray}
where $\hat\tau=\hat\mu_1-\hat\mu_0$,  $s_g^2$ is the sample variance in group $g$, $n^{-1}\hat{V}=\frac{s_1^2}{n_1}+ \frac{s_0^2}{n_0}$, $n^{-1}V=\frac{\sigma_1^2}{n_1}+ \frac{\sigma_0^2}{n_0}$,  $u=\frac{s_1^2/\sigma_1^2}{s_0^2/\sigma_0^2}\sim F(n_1-1,n_0-1)$, and
\begin{eqnarray*}
\begin{aligned}
 h^*(u) &=\sqrt{ \frac{(n-2) [u\sigma_1^2/n_1+\sigma_0^2/n_0] }{n^{-1}V\, [(n_1-1)u+n_0-1] } },\\
 f(u) & = \frac{\left[u\sigma_1^2/n_1 +\sigma_0^2/n_0\right]^2 }{u^{2}\sigma_1^4/[n_1^2(n_1-1)] +\sigma_0^4/[n_0^2(n_0-1)] }, \\
h(u) &=t_{f(u),1-\frac{\alpha}{2}} h^*(u).
\end{aligned}
\end{eqnarray*}
 The exact equivalence power (equation (A3) of  \cite{tang:2018a})  can be reexpressed as 
\begin{equation}\label{power_equi_tun}
P_{\text{equi}} =\int_0^\infty \left\{Q_{n-2}\left[-h(u),\delta_2; 0,R(u)\right]- Q_{n-2}\left[h(u), \delta_1;0,R(u)\right]\right\} dF_{n_1-1,n_0-1}(u) 
\end{equation}
where   $\delta_2=\frac{M_l-\tau_1}{\sqrt{n^{-1}V}}$, $\delta_1=\frac{M_u-\tau_1}{\sqrt{n^{-1}V}}$ and  $R(u)=\frac{\sqrt{n-2}\,(\delta_1-\delta_2)}{2h(u)}$. Please see \cite{tang:2018a} for numerical examples.

\section{ANCOVA}
\cite{tang:2018e, tang:2018a}  derived the exact power formulae for ANCOVA analysis of two-arm trials. 
Below we present  more general results for unstratified or stratified multi-arm randomized trials. Suppose subjects are randomized to $K^*=K+1$ treatment groups ($g=0,\ldots,K$) within each of $h$ strata. In an unstratified trial, we set $h=1$.  
Subjects in treatment group $g$  are modeled by
$$ y_{gi} = \mu_g + z_{gi_1}\alpha_1 +\ldots +z_{gi_{r-1}} \alpha_{r-1} + \xv_{gi}' \betav + \varepsilon_{gi} = \eta+ \delta_g + z_{gi_1}\alpha_1 +\ldots +z_{gi_{r-1}} \alpha_{r-1} + \xv_{gi}' \betav + \varepsilon_{gi} $$
where $z_{gi_k}$ ($k=1,\ldots,r-1$) is the  indicator variable for the pre-stratification factors, $\mu_g$ is the effect for treatment group $g$, $\xv_{gi}$ is the $q\times 1$ vector of baseline covariates, $\varepsilon_{gi}\sim N(0,\sigma^2)$, $\eta=\mu_0$ and $\delta_g =\mu_g-\mu_0$.
In general, $r$ equals the number of strata $h$. In trials with  multiple stratification factors, $r<h$ if there is no  interaction between some stratification factors. 
By the same arguments as the proof of equation (15) in  \cite{tang:2018e}, we obtain
the variance for the linear contrast with coefficients $(l_0,\ldots, l_K)'$   
$$\text{var}\left(\sum_{g=0}^K l_g \hat\mu_g\right) =\sigma^2 V_l\,\left(1+\frac{q}{n-q-r-K+1}\tilde{\Upsilon}\right)$$
where $\sum_{g=0}^K l_g=0$, $\bar{\zv}_g$ is the mean of $\zv_{gi}=(z_{gi_1},\ldots, z_{gi_{r-1}})'$ in group $g$,   $S_{zz}=\sum_{g=0}^K\sum_{i=1}^{n_g} (\zv_{gi}-\bar{\zv}_g)^{\otimes 2}$,
 $\tilde{\Upsilon}$ is a function of the covariate $\xv_{gi}$'s, and
$V_l= \sum_g l_g^2/n_g + (\sum_g l_g \bar{\zv}_g)'  S_{zz}^{-1}(\sum_g l_g \bar{\zv}_g)$. In a two arm trial  \citep{tang:2018e}, $V_l= \left[\sum_{s=1}^h \frac{n_{s1}n_{s0}}{n_{s1}+n_{s0}}\right]^{-1}$ if there is no restriction on the stratum effect  (i.e. $r=h$), where $n_{sg}$ is the number of subjects in stratum $s$, treatment group $g$.
A constant treatment allocation ratio  is commonly used in practice. Then $\bar{\zv}_0=\ldots=\bar{\zv}_K$ and $V_l= \sum_g l_g^2/n_g$.
Let $\tau_1=\sum_g l_g\mu_g$, $f=n-q-r-K$, and $f_2= f+1$.
When $\xv_{gi}$'s are normally distributed, $\tilde{\Upsilon} \sim F(q, f_2)$ and the exact power for the superior or NI test is
\begin{eqnarray}\label{power_ancova_un}
\begin{aligned}
P_{\text{sup/ni}} = \int_0^\infty \text{Pr}\left[ t\left(f,\sqrt{\frac{(\tau_1-M_0)^2} {\sigma^2V_l(1+q\tilde{\Upsilon}/f_2)}}\,\right) > t_{f,1-\frac{\alpha}{2}} \right] d F_{q,f_2}(\tilde{\Upsilon}).
\end{aligned}
\end{eqnarray}
Formula \eqref{power_ancova_un} also provides very accurate power estimate for nonnormal covariates \citep{tang:2018a}.
In equivalence trials, the exact power is
\begin{equation}\label{power_equi_ancova}
P_{\text{equi}} =\int_0^\infty \left\{Q_f\left[-t_{f,1-\alpha/2},\delta_2(\tilde{\Upsilon}); 0,R(\tilde{\Upsilon})\right]- Q_f\left[t_{f,1-\alpha/2}, \delta_1(\tilde{\Upsilon});0,R(\tilde{\Upsilon})\right]\right\} dF_{q,f_2}(\tilde{\Upsilon})  
\end{equation}
where $\delta_2(\tilde{\Upsilon})=\frac{M_l-\tau_1}{ \sqrt{\sigma^2 V_l(1+q\tilde{\Upsilon}/f_2)} }<0$, $\delta_1(\tilde{\Upsilon})=\frac{M_u-\tau_1}{\sqrt{\sigma^2 V_l(1+q\tilde{\Upsilon}/f_2)}}>0$ and  $R(\tilde{\Upsilon})=\frac{\sqrt{f}(\delta_1-\delta_2)}{2 \,t_{f,1-\alpha/2}}$.
The exact power formulae (equation (A1) of   \cite{tang:2018a}, equation (30) of \cite{tang:2018e}) for two arm trials are equivalent to equation \eqref{power_equi_ancova} at $K=1$.

The power formulae  \eqref{powertsw_equi},  \eqref{power_equi_tun}, \eqref{power_ancova_un} and \eqref{power_equi_ancova} are of the form $\int_0^\infty P_c( x) dF_{f_1,f_2}(x)$, and can be calculated as
\begin{equation}
 P= \int_0^\infty P_c( x) dF_{f_1,f_2}(x)  = \int_0^1 P_c\left[ F_{f_1,f_2}^{-1}(\nu)\right]\,d\nu.
\end{equation}

Below we give three  hypothetical examples. Sample R code is provided in the Supplementary Material. In each example, the simulated  (SIM) power is evaluated based on $4,000,000$ simulated datasets. There is more than $95\%$ chance
that the SIM power lies within $0.05\%$ of the true power.
 In example $1$, we perform  the power calculation for a superiority trial. Subjects are randomized equally into $K^*=3$ groups ($K=2$ experimental, or control treatment) stratified by gender ($z_{gi_1}=1$ for male, $0$ for female) and age  ($z_{gi_2}=1$ if old, $0$ otherwise).  There are $6$ subjects per treatment group per stratum ($n_0=n_1=n_2=24$, $n=72$).  
There is no interaction between age and gender ($r=3$, $h=4$), and the outcome is normally distributed as
  $$ y_{gi} \sim N\left[ \mu_g + 0.6 \,z_{gi_1} + 0.3 \,z_{gi_{2}} + 0.5\,x_{gi} , 1\right]$$
where $(\mu_0,\mu_1,\mu_2)=(0,0.6,0.9)$ and $x_{gi} \sim N(0.2 z_{gi_1}+0.4 z_{gi_2},1)$.
We compare each experimental treatment versus control treatment at the Bonferroni-adjusted one tailed  significance level of $\alpha/2=0.0125$. The exact power by formula \eqref{power_ancova_un}   is 
$78.63\%$ and $41.39\%$, and the SIM power is $78.62\%$ and $41.39\%$ respectively for the two tests. 

Example $2$ has similar setup to example $1$ except that $(\mu_0,\mu_1,\mu_2)=(0,0.05,0.1)$ and the sample size is $30$ per group per stratum ($n_0=n_1=n_2=120$, $n=360$).
The aim is to establish the equivalence of each experimental treatment versus control treatment at $\alpha/2=0.0125$. The margin is $(M_l,M_u)=(-0.5,0.5)$.  The exact power by formula \eqref{power_equi_ancova}   is 
$79.14\%$ and $86.72\%$ respectively for the two tests, while the SIM power is $79.14\%$ and $86.71\%$. 

In example $3$, we  design a three-arm ``gold standard'' NI trial \citep{pigeot:2003}. It consists of placebo ($g=0$), an active control treatment ($g=1$) and an experimental treatment ($g=2$). The set up 
is similar to example $1$ except that $(\mu_0,\mu_1,\mu_2)=(0,1,1.1)$, and the sample size is $10$ per group per stratum ($n_0=n_1=n_2=40$, $n=120$). Two tests are conducted  at the one-sided significance level of $\alpha/2=0.025$. 
Test $1$ evaluates the superiority of treatment $1$ over placebo. The  power for this test (exact $P_1=99.29\%$, SIM  $99.28\%$) is very close to $1$. 
In test $2$, we assess the noninferiority of treatment $2$ to treatment $1$ by demonstrating that treatment $2$ preserves at least $50\%$ of the efficacy of treatment $1$ compared to placebo
(i.e. $\frac{\mu_2-\mu_0}{\mu_1-\mu_0}>  50\%$ or  $\mu_2-0.5\mu_1-0.5\mu_0>0$). 
The exact power of test $2$  is $P_2=86.41\%$ (SIM power $86.41\%$).  The noninferiority is claimed only if both tests are significant \citep{pigeot:2003}, and the overall power is at least 
$P_1+P_2-1 =85.70\%$ while the simulated power is $85.80\%$.

\bibliographystyle{Chicago}

\bibliography{tsize_comment}

\begin{thebibliography}{}

\bibitem[\protect\citeauthoryear{Moser, Stevens, and Watts}{Moser
  et~al.}{1989}]{moser:1989}
Moser, B.~K., G.~R. Stevens, and C.~L. Watts (1989).
\newblock The two-sample t test versus {S}atterthwaite's approximate {F} test.
\newblock {\em Communications in Statistics -- Theory and Methods\/}~{\em 18},
  3963 -- 75.

\bibitem[\protect\citeauthoryear{Pigeot, Schafer, Rohmel, and Hauschke}{Pigeot
  et~al.}{2003}]{pigeot:2003}
Pigeot, I., J.~Schafer, J.~Rohmel, and D.~Hauschke (2003).
\newblock Assessing non-inferiority of a new treatment in a three-arm clinical
  trial including a placebo.
\newblock {\em Statistics in Medicine\/}~{\em 22}, 883 -- 99.

\bibitem[\protect\citeauthoryear{Tang}{Tang}{2018a}]{tang:2018e}
Tang, Y. (2018a).
\newblock Exact and approximate power and sample size calculations for analysis
  of covariance in randomized clinical trials with or without stratification.
\newblock {\em Statistics in Biopharmaceutical Research\/}~{\em 10}, 274 --
  286.

\bibitem[\protect\citeauthoryear{Tang}{Tang}{2018b}]{tang:2018a}
Tang, Y. (2018b).
\newblock A noniterative sample size procedure for tests based on t
  distributions.
\newblock {\em Statistics in Medicine\/}~{\em 37}, 3197 -- 213.

\end{thebibliography}

\end{document}